\documentclass{ifacconf}

\usepackage{graphicx}      
\usepackage{natbib}        
\usepackage{changes} 
\usepackage{xcolor}
\begin{document}
\begin{frontmatter}

\title{An Agent Model Abstraction for Human–AI Teaming Cognitive Coupling\thanksref{footnoteinfo}} 

\thanks[footnoteinfo]{This work was supported by the Horizon Europe research and innovation program under Grant Agreement No 101135990 (AI4Work). \copyright~2026 the authors. This work has been accepted to IFAC for publication under a Creative Commons Licence CC-BY-NC-ND.}

\author[X]{Kolitha Kottagaha W.M} 
\author[X]{Jos A.C. Bokhorst} 
\author[Y,Z]{Ben Gaffinet} 
\author[X]{Christos Emmanouilidis} 

\address[X]{University of Groningen, 9700 AV, Groningen, The Netherlands (e-mail: (kolitha.kottagaha; j.a.c.bokhorst; c.emmanouilidis)@rug.nl).}
\address[Y]{Luxembourg Institute of Science and Technology, Belval, Luxembourg (e-mail: ben.gaffinet@list.lu).}
\address[Z]{Université de Lorraine, CNRS, CRAN, Nancy, France}

\begin{abstract}                
Industrial environments increasingly rely on collaboration between humans and AI-enabled agents. Effective teamwork requires aligning how agents perceive situations, plan actions to pursue goals, and adapt to changing conditions, yet existing systems lack mechanisms for cross-agent cognitive processes coupling. This paper presents a conceptual cognitive agent model that formalises cognitive coupling through eight components: Input, Process, Output, State, Value, Memory, World Model, and Goal. The model abstracts how agents coordinate and co-regulate their cognitive cycles, providing a basis for analysing distributed cognition and designing cognitively interoperable human–AI systems.

\end{abstract}


\end{frontmatter}

\section{Introduction}

Advances in Artificial Intelligence (AI) have shifted collaborative working environments from humans using AI systems as passive tools to humans and AI working together as team members within complex socio-technical systems. In modern cyber-physical and industrial environments, cognitive capabilities such as perception, reasoning, and learning are increasingly distributed across both human and AI-enabled agents who collaborate towards shared goals. Effective teaming in such settings, therefore, depends on the ability of these different types of agents to perceive, reason, and act in alignment with common objectives while continuously adapting to changing operational and environmental conditions. Achieving such coordination requires cognitive interoperability, that is, the ability of human and artificial agents to align their thoughts and perceptions of information, enabling mutual understanding and shared intentions \citep{Jana2025}. Traditional models of human-automation interaction are often conceptualised as dyadic and task-oriented, emphasising task allocation and supervisory control rather than cognitive collaboration \citep{Naikar2023}. However, Industry 5.0 and cyber-physical human systems demand models capable of abstracting how cognitive processes across human and AI entities align to sustain adaptive collaboration \citep{Leito2022,Jana2025}.

Cognitive science and systems engineering provide complementary perspectives for addressing this challenge. From a cognitive perspective, theories of distributed and joint cognition describe how cognitive functions extend beyond the individual through interaction and coordination \citep{Naikar2023}. In parallel, research on team cognition demonstrates how shared mental models and communication dynamics support effective collective performance \citep{Cooke2024, Canonico2019}. Within system engineering, multi-agent and cyber–physical systems provide the structural basis for embedding these ideas, in which agents, both human and artificial, can exhibit adaptive and goal-directed behaviour through decentralised interactions \citep{Leito2022}. Cognitive architectures such as ACT-R, LIDA, ICARUS, and SOAR further advance this understanding by modelling how agents perceive, reason, and learn \citep{Franklin2014, Laird2012, Anderson1993,Langley2006}. The next step, however, is to extend such architecture to represent the coupling processes that enable human and artificial agents to coordinate and adapt within shared environments.

Cognitive coupling is understood here as the alignment between the cognitive characteristics of human and artificial agents. The proposed model formalises this coupling as a mechanism to achieve cognitive interoperability between human and AI-enabled agents in collaborative environments. Each agent is represented as a cognitive unit composed of eight components: Input, Process, Output, State, Value, Memory, World Model, and Goal. The agent interacts with the environment while maintaining alignment of its intentions and actions with those of other agents through a Shared Goal Space that represents common objectives. The proposed model provides a foundation for analysing and designing cognitively interoperable human-AI teaming. The remainder of this paper introduces the design methodology, theoretical grounding, describes the model components and their relationships, and discusses its implications.

\section{Background and Related Work}

Research in team cognition has recognised that effective teamwork depends not only on task coordination but also on the sharing and integration of cognitive processes among members. Studies on shared mental models and collective intelligence highlight that communication and mutual awareness enable teams to function as a coherent cognitive system rather than as a collection of individuals \citep{Cooke2024, Canonico2019}. These insights apply not only to human teams but also to human–non-human teaming, where AI-enabled agents demonstrate cognitive capabilities that support collaborative work. 
Building on this perspective, the joint and distributed cognition studies provide a broader view of how cognitive functions extend beyond individuals through interaction with other agents and artefacts. Joint cognitive systems can be seen as integrated human–technology units that interpret situations, manage complexity, and act collectively within dynamic environments \cite{Hollnagel2005}. Similarly, cognition in modern human-AI systems can emerge from coordinated activities across agents \cite{Naikar2023}. In such settings, agents continuously align their perceptions and adapt their actions in response to dynamic conditions within the system. Together, these perspectives highlight cognition as both an individual capability and a phenomenon that expands through the interactions of collaborating agents.

\begin{figure}
    \centering
    \includegraphics[width=0.8\linewidth]{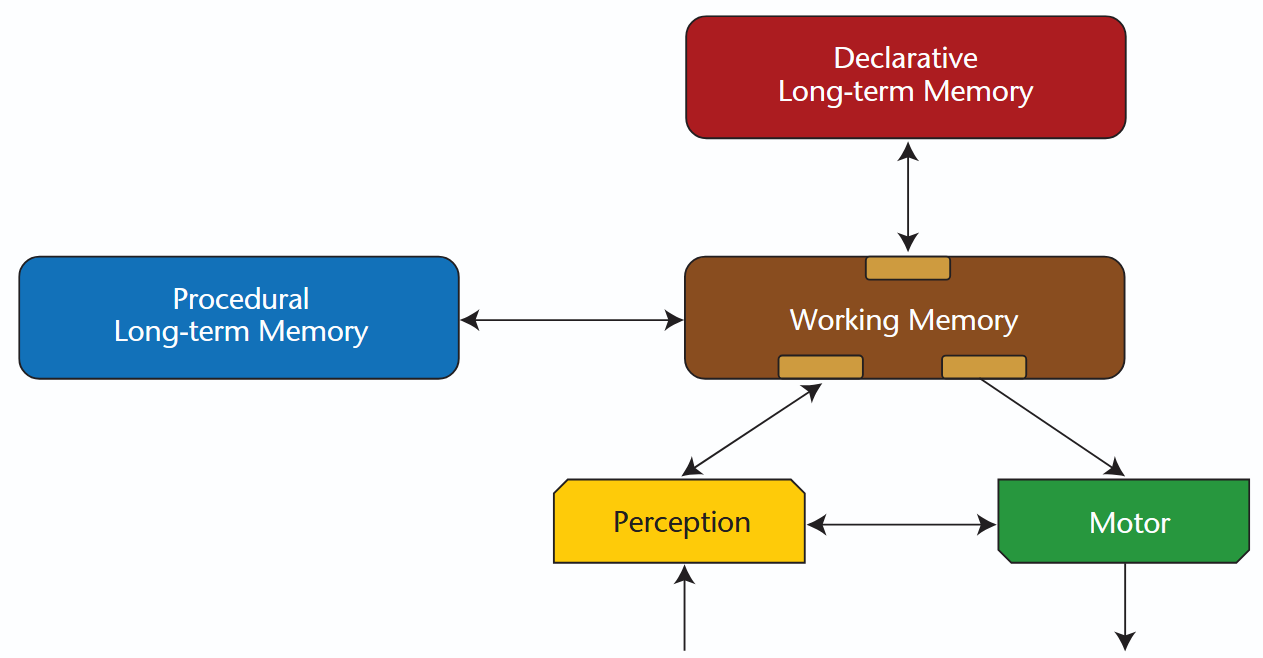}
    \caption{Common model of cognition,  \cite{laird2017standard}.}
    \label{fig:common-model-of-cognition}
\end{figure}

Complementing these theoretical perspectives, research in cognitive and cyber–physical systems has explored how individual agents can perceive, reason, and act within complex environments. Cognition can be modelled via cognitive architectures, abstracting the function of agents with human-like behaviour. Architectures such as ACT-R, SOAR, ICARUS, and LIDA formalise the internal cognitive cycle of an agent, how it processes inputs, retrieves knowledge, forms intentions, and selects actions \citep{Anderson1993, Laird2012, Franklin2014, Langley2006}.  These architectures offer a theoretical grounding for the invariants of cognition (i.e. structure) and computational models that follow said theory. For example, the common model of cognition by \cite{laird2017standard} offers a high-level standard model that abstracts a cognitive agent (see figure \ref{fig:common-model-of-cognition}). In parallel, studies on distributed control and collective intelligence in industrial cyber-physical systems demonstrate how multiple autonomous agents can coordinate at the system level to achieve robustness, adaptability, and shared autonomy \citep{Leito2022}. The importance of a cognitive-level of integration and coupling between humans and technical entities has also been highlighted in the context of human-in-the-loop of cyber-physical systems \citep{Emmanouilidis2019}. Cognitive functions, such as perception and situation awareness, recognition and categorisation, memory and reflection learning, reasoning and belief, interaction and decision-making, problem solving and planning, monitoring and prediction, choice and decision-making, all the way to actions taken and executed, were proposed as dimensions of such cognitive coupling. Recent work on cognitive digital twins further highlights the need for cognitive interoperability, where human and artificial agents must interpret goals, decisions, and context in compatible ways to sustain collaboration \citep{Jana2025}. In addition, decisional interoperability emphasises that human-AI teams require shared decision spaces and compatible interpretative frameworks to support joint reasoning and coordinated action \citep{Emmanouilidis2025}. 

Existing literature provides important foundations for understanding collective work, individual cognition, and system-level coordination in human-AI settings. What remains less specified is a representation that captures both the internal organisation of cognitive processes of human and artificial agents and the way their cognitive cycles become coupled during collaborative work. Existing work often treats team-level coordination, individual cognition, and system architecture separately, which makes it difficult to trace how goals, evaluations, and actions flow across agents. The next section introduces a conceptual cognitive agent model that combines internal cognitive components with a Shared Goal Space and explicit interaction links between agents and their environment. This model is intended as a step toward making the mechanisms of cognitive coupling in human–AI teaming explicit.

\section{Design Methodology}

This work introduces  a conceptual abstraction of the cognitive coupling needed in human-AI teaming. It connects the internal cognitive mechanisms of human and artificial agents with the interaction processes that couples them during collaborative work. The design starts by identifying recurring cognitive functions in the cognitive architectures literature. These functions are organised into a generic cognitive unit that can abstract both human and artificial agents at a functional level. The model focuses on how each agent organises its internal cognitive processes and how these processes become connected with those of other agents within a coupled cognitive system. These cognitive units are extended with system-level structures required to represent collaboration among agents. The model is represented via SysML block diagrams. SysML is appropriate for the representation of system structure, component relations, and information flow within an integrated system. In this model, the block definition diagram describes the main elements of the cognitive system, the internal block diagram describes the flow of information within an agent, and the system-level internal block diagram shows how information flows among agents and other shared components. The model was validated at a conceptual level. This involved checking the consistency of the proposed components with recurring functions in cognitive architecture literature, the coherence of the internal cognitive cycle, and the ability of the system-level structures to represent agent coupling.

\section{Model Description}

\begin{figure}[ht]
    \centering
    \includegraphics[width=0.7\linewidth]{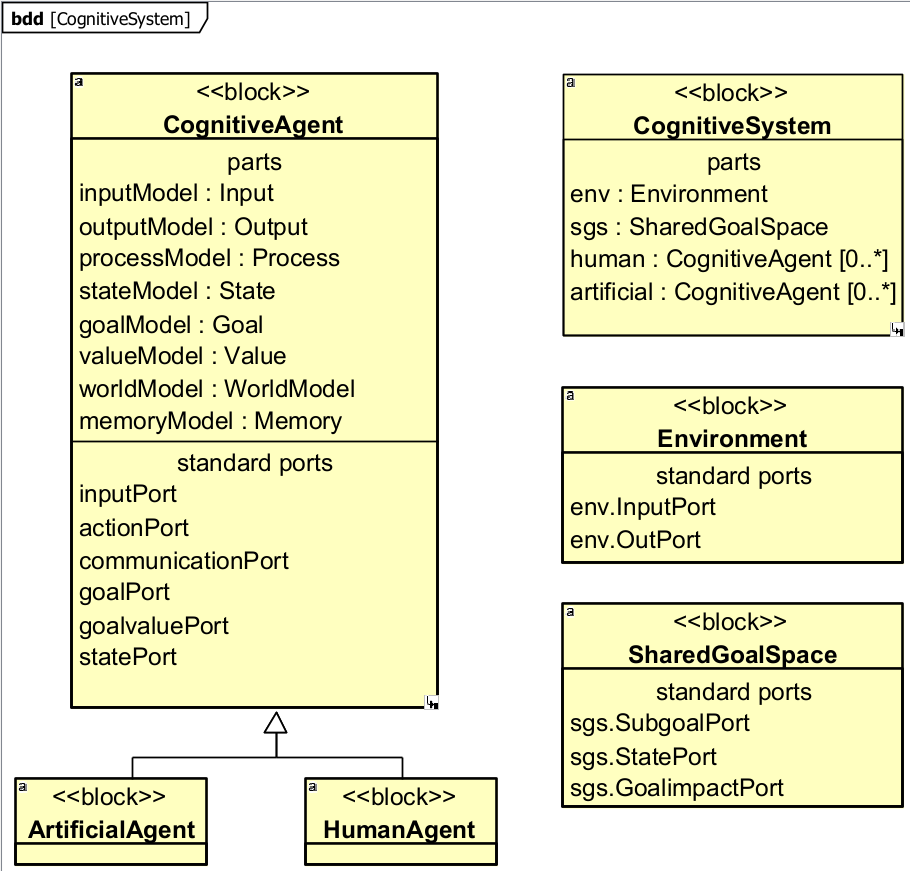}
    \caption{Block Definition Diagram}
    \label{fig:BDD}
\end{figure}

\begin{figure*}
    \centering
    \includegraphics[width=0.8\linewidth]{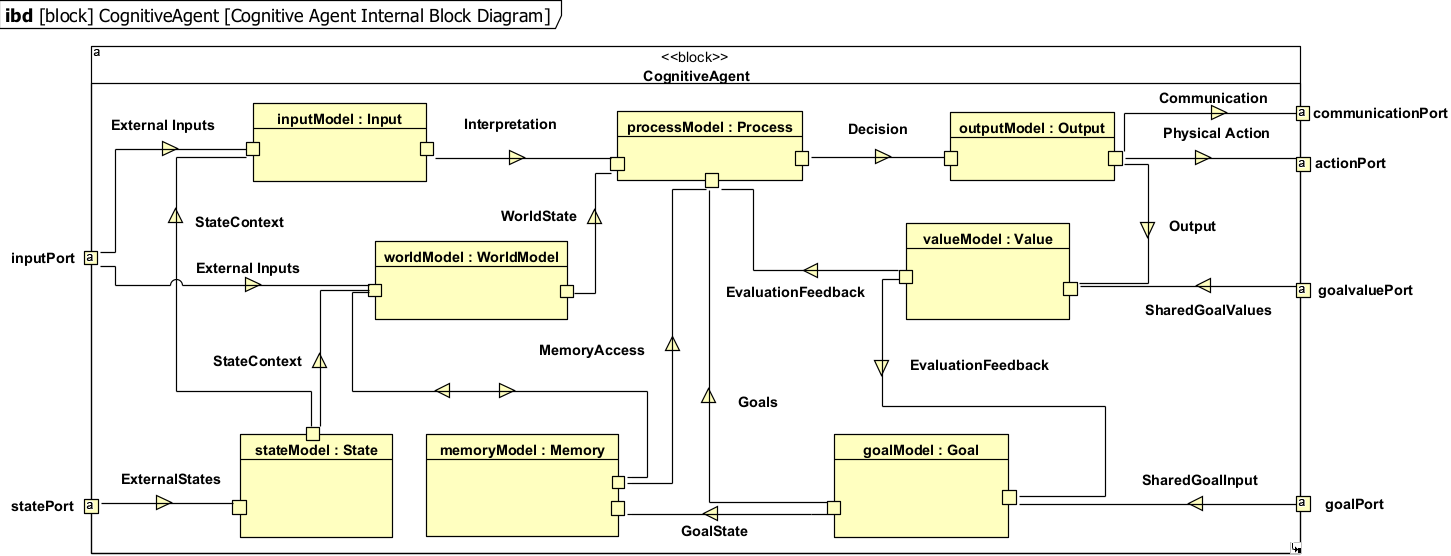}
    \caption{Internal block diagram of a cognitive agent}
    \label{fig:cognitive-agent}
\end{figure*}

The proposed model conceptualises human and AI-enabled agents as cognitively capable units that operate within a shared environment and coordinate their activity through a shared goal representation. The Cognitive System integrates Cognitive Agents with  external \textit{Environment} representations and \textit{Shared Goal Space}. The \textit{Environment} offers sensory and task context. Within it, agents perceive, act, and receive feedback, while the \textit{Shared Goal Space} maps collective goals and maintains a common representation of goal states and performance information. Cognitive coupling emerges as agents interact through these structures, aligning their intentions through the \textit{Shared Goal Space}, coordinating actions through communication, and updating their internal states and goals in response to shared feedback. Figure~\ref{fig:BDD} shows a \textit{CognitiveAgent} composed of eight interrelated components, which are \textit{Input}, \textit{Process}, \textit{Output}, \textit{State}, \textit{Value},\textit{Memory}, \textit{WorldModel} and \textit{Goal}. This organisation follows common patterns from cognitive architectures that integrate perception, memory, learning, and goal management \citep{Kotseruba2020}. Components capture core cognitive functions required for perception, interpretation, decision-making, evaluation, and adaptive goal regulation. The Cognitive Agent is instantiated in two  forms: a \textit{HumanAgent} and an \textit{ArtificialAgent}. These share the same cognitive structure but differ on how their functions are realised. 

\subsection{Model Components}

\noindent\textbf{\textit{Input:}} The \textit{Input} component defines how an agent perceives and selects information from the environment, other agents, and its own internal states. It structures the perceptual process of receiving sensory and communicative signals and interpreting them into representations for use within the agent’s internal mechanisms. As part of this, the \textit{Input} includes attentional functions that determine which information to prioritise, filter, or suppress. These mechanisms direct the agent’s perceptual focus toward information relevant to its current state, world model, or goals, so that only relevant inputs are handled in further processing. In collaborative environments, the \textit{Input} supports perception- and communication-driven updates, enabling agents to maintain shared context and interpret changes in their environment, goal states, and other agents' activities.

\noindent\textbf{\textit{Process:}} The \textit{Process} component formalises the internal cognitive mechanisms that transform input into decisions.  It does not describe a single algorithm but instead abstracts a set of sub-processes, such as situation interpretation, reasoning, option evaluation, and action selection. Through these sub-processes, perceptual information from the \textit{Input} is combined with evaluative feedback from the \textit{Value} and contextual information from the \textit{State}, \textit{World Model}, \textit{Memory}, and \textit{Goal}. The \textit{Process} therefore determines how the current situation is understood, which actions are relevant, and which behaviour best supports the agent's local goals and the shared goals of the system.

\noindent\textbf{\textit{Output:}} The \textit{Output} component defines how an agent expresses and executes its decisions. It receives selected actions from the \textit{Process} component and translates them into effects that influence the environment, update the \textit{SharedGoalSpace}, or interact with other agents. These outputs may involve physical actions that affect the external entities or communicative acts that convey intentions, progress updates, or coordination signals. In collaborative settings, the \textit{Output} component plays a critical role in maintaining transparency and coordination, as its communicative outputs allow other agents to interpret ongoing activities and adjust their own behaviour accordingly.

\noindent\textbf{\textit{Value:}}The \textit{Value} component defines the evaluation criteria an agent uses to assess the outcomes of its actions, and the progress towards local and shared goals. It receives information on recent actions from the \textit{Output}. Through this process, the \textit{Value} component determines whether selected actions contributed to or hindered the agent's progress towards high level objectives and whether adjustments to behaviour or goals are required. The component produces evaluative feedback that is passed to the \textit{Process} and \textit{Goal} components, influencing decisions, updating priorities, and supporting goal revision when necessary.  The \textit{Value} incorporates multiple evaluative factors depending on the agent and goal context, such as efficiency, safety, workload, and alignment with shared goals. By synthesising these considerations into an evaluative feedback, the \textit{Value} component supports adaptive behaviour and maintains coordination among agents engaged in collaborative activities.

\noindent\textbf{\textit{Goal:}}The \textit{Goal} defines the agent’s local goal hierarchy. It contains both the higher-level local goals the agent contributes to and the sub-goals required to achieve them. Local goals correspond to system-level goals derived from the \textit{SharedGoalSpace}. They can also arise from the agent’s interpretation of new requirements or blocked progress. Goal generation defines how an agent establishes or revises goals in response to shared goals, feedback, or situational changes, deciding what to achieve next and how to proceed. Achieving a local goal requires intermediate steps, sub-goals that specify actions or conditions needed for progress. If accessing an out-of-reach item is a goal, a sub-goal is to move the agent to a position such that the item becomes reachable. Local goals and sub-goals are updated in response to feedback from the \textit{Value} component and context from the \textit{State} and \textit{World Model}. When conditions change or progress stalls, the agent may revise goals, add sub-goals, or shift focus. Managing goal formation and decomposition enables flexibility when demands change.

\noindent\textbf{\textit{Memory:}}The \textit{Memory} component stores information beyond the immediate perception-action cycle. It encompasses forms of memory that support different cognitive functions: episodic memory for event-specific experiences, semantic or declarative memory for general knowledge, procedural memory for learned skills and action patterns, and working memory for maintaining task-relevant information over short intervals. \textit{Memory} interacts with other components. The \textit{Process} component uses \textit{Memory} to reason about past situations or adapt to new constraints. The \textit{World Model} stores and retrieves elements via Memory for coherence over time. Memory also helps the \textit{Goal} component recall past goals and sub-goals, enabling reuse and adaptation for continuity in planning and coordination.

\noindent\textbf{\textit{World Model:}} The \textit{World Model} is the agent’s internal representation of the external environment and other agents. It captures aspects of the situation that the agent maintains internally, forming a representation of the surrounding context. Unlike the Environment, which denotes the external world itself, the \textit{World Model} is the agent’s internalised interpretation of that world and functions as a minimum viable representation that supports decision-making. It also enables the agent to anticipate how the situation may unfold based on current information, past experience, and possible actions. Information arriving from the environment and other agents is integrated with the \textit{World Model}, allowing the agent to detect changes, updates, or inconsistencies in the situation. As actions are taken or new evidence becomes available, this representation is revised to maintain continuity with the unfolding task context. The \textit{Process} component consults the World Model when selecting actions, ensuring that decisions are informed by the agent’s understanding of its surroundings.

\noindent\textbf{\textit{State:}}The \textit{State} component represents the agent’s internal condition. It includes self-related information such as the agent’s current capabilities, readiness, workload, confidence, and other factors that determine the agent's action possibilities at a given moment. These internal conditions influence how new information is interpreted and shape which actions are considered feasible or appropriate within the current situation. In addition to its own condition, the agent’s state can include internally maintained information about other agents when relevant to coordination. Such information may stem from communication or observation and becomes part of the agent’s internally represented state. These states support smoother collaboration by enabling agents to anticipate each other’s availability or engagement and to adjust their actions accordingly.

\noindent\textbf{\textit{Shared Goal Space:}}The \textit{SharedGoalSpace} represents the common objective structure that guides coordination between agents. It provides a centralised space for maintaining system-level goals, constraints, and task states. Through this shared representation, agents retrieve objectives and the evolving demands of the system goals. The \textit{SharedGoalSpace} ensures that all agents interpret high-level goals consistently, forming a reference point for reasoning, decision-making, and sub-goal formation. A core function of the \textit{SharedGoalSpace} is to support \textit{goal decomposition}. High-level system goals are interpreted and distributed, from which each agent derives its own sub-goals. This relationship enables agents to pursue local objectives while maintaining alignment with system goals. 

\subsection{Cognitive Coupling Between Agents}
\begin{figure*}
    \centering
    \includegraphics[width=0.7\linewidth]{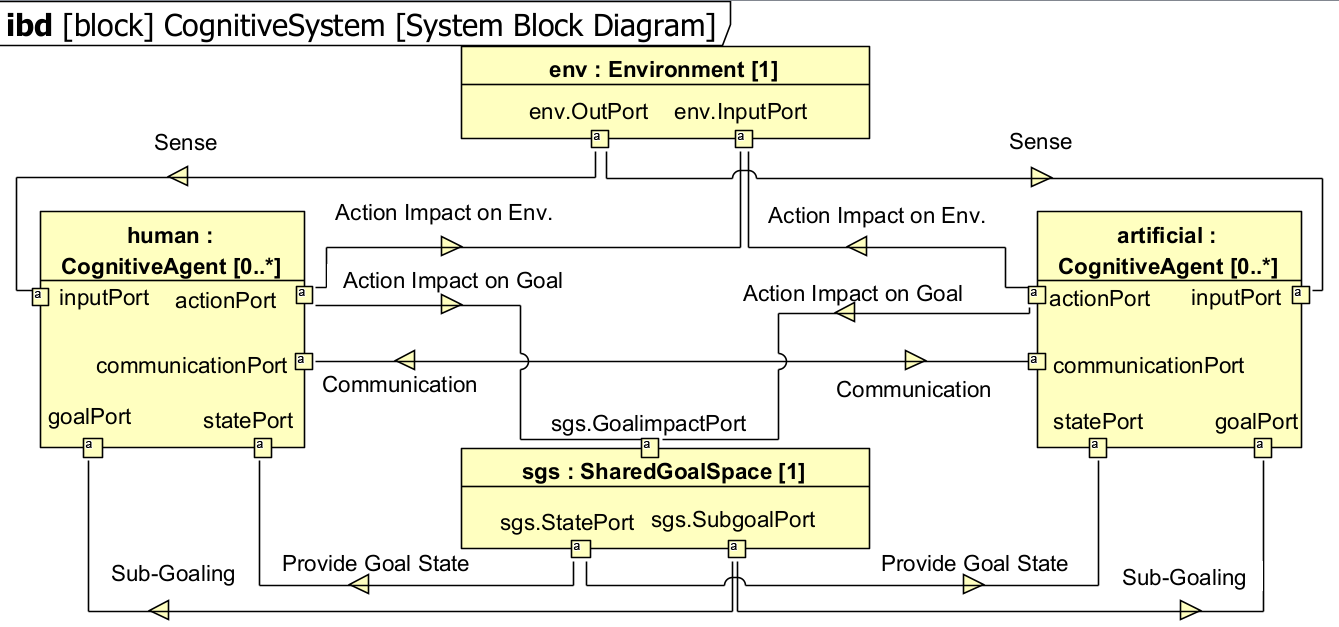}
    \caption{Internal Block Diagram of the Cognitive System}
    \label{fig:IBD_System}
\end{figure*}
In the proposed architecture, cognitive coupling refers to the way in which the internal cognitive processes of one agent become relevant for, and constrained by, those of other agents through structured interactions with the environment, other agents, and the \textit{SharedGoalSpace}. Figure~\ref{fig:IBD_System} shows that human and artificial agents are not isolated decision-makers. Their cognition is linked through three main mechanisms. These are shared perception–action loops with the environment, exchanges of goal-related information via the \textit{SharedGoalSpace}, and direct communication between agents. First, actions performed by one agent modify the environment, thereby changing what other agents perceive. This perception–action coupling means that an agent’s Output directly shapes the Input of others, creating a dependency between their internal cognitive cycles. Second, both agents read from and write to the \textit{SharedGoalSpace}. Goal-state information from the \textit{SharedGoalSpace} constrains sub-goal formation within each agent, while sub-goal contributions and action impacts update the shared representation of progress. This establishes a link between individual intentions and collective objectives. Third, state-related and communication signals allow agents to exchange information about their status or active task, providing additional constraints and opportunities for coordination. These mechanisms ensure that the agent’s cognition not only operates independently but what each agent perceives, intends, and decides is shaped by the evolving actions, states, and goals of other agents within the system. In this model, cognitive coupling results from interaction pathways that connect internal cognitive components across agents.

\section{Relation to Cognitive Architectures}

\subsection{Cognitive Functions}
Research on cognitive architectures over the past decades has converged on a set of recurring functional categories, including perception, attention, memory, learning, reasoning,, and action selection \citep{Kotseruba2020}. Examining the proposed model alongside functional categories reveals how its organisation resonates with established architectures and where it expands them to support collaboration between human and artificial agents.

For perception, the \texttt{Input} component matches the perceptual layers in many architectures. It handles sensory and communicative signals, selecting and structuring information before central processing. This aligns with ACT-R, Soar, and LIDA, which separate input from central processing and remain agnostic to perceptual algorithms. Attention is handled by mechanisms that select or filter information. In the proposed model, \texttt{Input} filters or prioritises data, while \texttt{Process} chooses which perceived elements enter decision-making based on goals and feedback. Reasoning and action selection correspond to \texttt{Process} and \texttt{Output}. \texttt{Process} integrates perception, knowledge, state, and values to generate actions. \texttt{Output} executes them and updates the environment and Shared Goal Space. This separation mirrors many architectures and highlights coordination functions vital for human–AI teams.

Existing architectures divide memory into episodic, semantic, and working forms. The proposed model assigns roles to different components: \texttt{Memory} provides long-term knowledge and episodic traces; \texttt{World Model} captures situational understanding, including representations of other agents and the environment;  \texttt{State} maintains transient internal conditions. Learning adapts internal representations based on experience. In the model, learning results from interactions between \texttt{Process}, \texttt{Output}, and \texttt{Value}. \texttt{Process} makes decisions; \texttt{Output} enacts them; \texttt{Value} evaluates outcomes against goals, producing feedback to update decisions, knowledge, and goals.

\subsection{The Common Model of Cognition}
The Common Model of Cognition proposed by \cite{laird2017standard} provides a high-level structure for human cognitive processes that can be related to the components of the proposed  architecture. The agents in the proposed architecture can either be artificial or human in nature. Artificial agents can be designed to follow the proposed cognitive agent architecture, but humans are natural systems whose design can't be altered. The internal cognitive state and processes of humans are not fully measurable and can only be partially inferred from behaviour.

A model that approximates a human agent using a cognitive architecture is compatible with the input/output behaviour of the proposed cognitive system architecture. Within the proposed architecture, agents receive four different types of inputs; \texttt{Sensing}, \texttt{Communication}, \texttt{State} and \texttt{Goal}. For artificial agents all the inputs are received in digital form, for instance data from sensors cover the \texttt{Sensing} input and \texttt{Goal} information is sent by an external entity that manages the \texttt{cognitiveSystem.SharedGoalSpace}. On the other hand, human agents either sense the environment directly with their senses or receive messages through Human-Machine Interfaces. In any case, external information is received by the senses and are processed by the \texttt{laird.Perception} block which corresponds to the \texttt{cognitiveAgent.Input} (see figure \ref{fig:common-model-of-cognition}). For cognitive architectures, perception is commonly modelled with multiple separate modules, one for each modality. As output, the \texttt{laird.Perception} block yields symbolic structures and are subject to attention that limits the quantity of symbolic information it can output. The \texttt{laird.Motor} block converts internal symbolic representations into external actions, mapping onto the \texttt{cognitiveAgent.Output} block.

The internal processing in cognitive architectures is driven by a central \texttt{laird.WorkingMemory} block that accesses all buffers and procedural rules that are stored in \texttt{laird.ProceduralMemory}. Rules take the form of conditions and actions and are, in principle, learnable. Declarative information is stored in \texttt{laird.DeclarativeMemory} where it can be reinforced and decayed based on the situations the modelled human encounters. The high-level structure of the common model remains flexible and can be adapted when implementing. This flexibility allows elements of the proposed model to be related to the Common Model at the level of cognitive function. For instance, the Common Model does not include a separate \texttt{WorldModel} component. Instead, the proposed model separates the agent’s situational representation as \texttt{cognitiveAgent.WorldModel}, while in the Common Model this role is mainly handled within \texttt{laird.WorkingMemory}. It maintains the agent's current understanding of the environment, task context, and other agents, and provides the situational information that the Process component uses to reason and select actions.


\section{Conclusion and Future Work}

The model introduced in the present work abstracts how human and artificial agents cognitively coordinate their action in shared tasks. It describes agents using eight internal components and connects them to the Environment and \textit{SharedGoalSpace}, clarifying how agents' decisions rely on shared context and other agents’ decisions and actions. The \textit{SharedGoalSpace} provides a common reference for collective objectives. By representing shared goals, the model demonstrates how agents remain aligned as tasks and situations change. It clarifies how agents influence one another through perception–action relations, communication, and shared goal updates, thereby creating coupling effects. While the model captures key cognitive and interaction processes, it remains a conceptual representation rather than an operational system. It does not describe specific algorithms or behavioural policies, nor does it address domain-specific constraints. The modelling challenges to represent a human individual's cognition in a computerised system are substantial. Cognitive architectures have found success in modelling human behaviour for diverse tasks and reproduce performance levels that closely match humans. To capture an individual's cognition the model needs to account for interpersonal differences such as working memory capacity. Validating models is a separate challenge as the modelled internal cognitive states are not observable, but only the resulting decisions and actions. 
Future work will explore how components aligned with this representation can be implemented and used to analyse performance in simulated or real collaborative tasks. The model also offers a conceptual basis for developing cognitive digital twins. Such twins aim to represent not only the physical or operational aspects of agents, but also the cognitive structures that support how agents perceive situations and select actions. By specifying how inputs, memory, world representations, values, and goals are organised and coupled across agents, the proposed model highlights the internal structure and coupling mechanisms that can guide the design of digital twins.
\vspace{-0.5em}
\bibliography{ifacconf}       

\end{document}